\documentclass{emulateapj}
\usepackage{multirow}
\usepackage{subfigure}

\def\xmm{\textit{XMM-Newton}}
\tabletypesize{\tiny}
\shortauthors{Lin et al.}
\begin{document}

\title{Discovery of Three
  Candidate Magnetar-powered Fast X-ray Transients from Chandra
  Archival Data}
\author{Dacheng Lin\altaffilmark{1,2}, Jimmy A. Irwin\altaffilmark{3},
Edo Berger\altaffilmark{4}, Ronny Nguyen\altaffilmark{2}}
\altaffiltext{1}{Department of Physics, Northeastern University, Boston, MA 02115-5000, USA, email: d.lin@northeastern.edu}
\altaffiltext{2}{Space Science Center, University of New Hampshire, Durham, NH 03824, USA}
\altaffiltext{3}{Department of Physics and Astronomy, University of Alabama, Box 870324, Tuscaloosa, AL 35487, USA}
\altaffiltext{4}{Harvard-Smithsonian Center for Astrophysics, 60 Garden Street, Cambridge, MA 02139, USA}

\begin{abstract}
It was proposed that a remnant stable magnetar could be formed in a
  binary neutron-star merger, leading to a fast X-ray transient (FXT) that
  can last for thousands of seconds.  Recently, Xue et al. suggested
  that CDF-S XT2 was exactly such a kind of source. If confirmed, such
  emission can be used to search for electromagnetic counterparts to
  gravitational wave events from binary neutron-star mergers that have
  short gamma-ray bursts and the corresponding afterglows seen off-axis and thus too weak to be
  detected. Here we report the discovery of three new FXTs, XRT
  170901, XRT 030511, and XRT 110919, from
  preliminary search over Chandra archival
  data. Similar to CDF-S XT2, these new FXTs had a very fast rise (less
  than a few ten seconds) and a plateau of X-ray flux
  $\sim$$1.0\times10^{-12}$ erg s$^{-1}$ cm$^{-2}$ lasting for 1--2 ks, followed by a steep
  decay. Their optical/IR counterparts, if present, are very weak,
  arguing against a stellar flare origin for these FXTs. For XRT
  170901, we identified a faint host galaxy with the source at the
  outskirts, very similar to CDF-S XT2. Therefore, our newly discovered
  FXTs are also strong candidates for  magnetar-powered X-ray transients
  resulting from binary neutron star mergers.

\end{abstract}

\section{INTRODUCTION}
\label{sec:intro}

The detection of both the gravitational wave (GW) and multiwavelength
electromagnetic (EM) signals from the binary neutron star (BNS) merger
event GW170817 marked the arrival of the multi-messenger era
\citep{ababab2017a}. The identification of EM counterparts is
essential to confirm the astrophysical origin of the GW events and to
advance our understanding of the physics of compact object
mergers. The EM emission from GW events depends on the merger types:
BNS, binary black hole (BBH), or neutron star (NS)-black hole
(BH). For BNS mergers, one expects the presence of short gamma-ray
bursts (sGRBs), afterglows, and kilonovas \citep[see][for a
review]{be2014}. There is evidence that the jets that produce sGRBs
and afterglows are collimated
\citep[e.g.,][]{bugrca2006,sobeka2006,fobeme2014}. This implies that a
majority of GW events are expected to be observed with sGRBs seen
off-axis, resulting in prompt emission and afterglows too weak to be detected. Kilonovae from BNS mergers can have large solid
angles but they are expected to be weak and hard to detect too
\citep{mebe2012,smchje2017}.

The EM emission from GW events also depends on the nature of the
post-merger remnant. It was proposed that some BNS mergers might produce a
supramassive, highly magnetized, and rapidly spinning NS (i.e.,
magnetar) rather than a BH. In this case, Zhang (2013) argued that the
afterglow powered by a rapidly spinning massive NS has a much wider
solid angle than the solid angle of the sGRB jet, so that gamma-ray
bursts (GRB)-less GW events can also have bright afterglows from a
dissipating proto-magnetar wind with a large solid angle. Such
afterglows can last for thousands of seconds. If this is the case,
such unique signals can be used to search for the EM counterparts to
GW events with jets viewed off-axis and can be used to probe massive
millisecond magnetars and stiff equation-of-state for nuclear matter \citep{gadiwu2013}.
 
Recently, \citet[][Xue19 hereafter]{xuzhli2019} reported the discovery of a very special
fast X-ray transient (FXT) from the \emph{Chandra} Deep
Field-South (CDF-S) Survey, CDF-S XT2. The FXT exhibited a fast rise ($<45$ s) to a
plateau that last about 2 ks, followed by a steep decay ($t^{-2}$). It
was not associated with a known GRB.  It lies at the outskirts of a
star-forming galaxy, as often seen in sGRBs but not in the long GRBs
(lGRBs).  Based on the properties of the light curve, spectral
evolution, host galaxy, location, and event rate, \citet{xuzhli2019}
argued CDF-S XT2 to be powered by a remnant stable magnetar from a BNS
merger. The magnetar was inferred to have a magnetic field order of
$10^{15}$ G and a spin period order of 1 ms \citep[Xue19;][]{xizhda2019, sulizh2019}. 

There is another FXT also discovered in the CDF-S Survey, i.e., CDF-S XT1 \citep{batrsc2017}. It rose within 100
s to the peak and then decayed approximately as $t^{-1.5}$ without a
clear presence of a plateau in the peak like CDF-S XT2. It was 
also argued to be magnetar-powered, but viewed through ejecta
\citep{sulizh2019}. We note that for this FXT, which has a very faint
host galaxy with unknown redshift, a very-high-$z$ GRB explanation
cannot be ruled out.

It is important to discover more similar FXTs in order to confirm their
astrophysical origin. We have been searching for important sources
from \emph{Chandra} X-ray archival data and have found three new FXTs very similar to CDF-S
XT2 in many aspects\footnote{During the preparation for this paper,
  we have discovered a new FXT \citep{liirbe2021atell}, which will be reported in a
  separate paper.}.  These three FXTs are denoted as X-Ray Transient
(XRT) 170901, XRT 030511, and XRT 110919, following the dates on which
they occurred. In Section~\ref{sec:reduction}, we describe the
data analysis of X-ray and other multiwavelength data.
In Section~\ref{sec:res}, we present the results of our analysis of the light
curves and X-ray spectra of the FXTs and our search for their host galaxies.
Our discussion of the nature of our sources is given in Section~\ref{sec:discussion} and  the conclusions of
our study are given in Section~\ref{sec:conclusions}.

\begin{figure}

\begin{center}
 \includegraphics[width=3.4in]{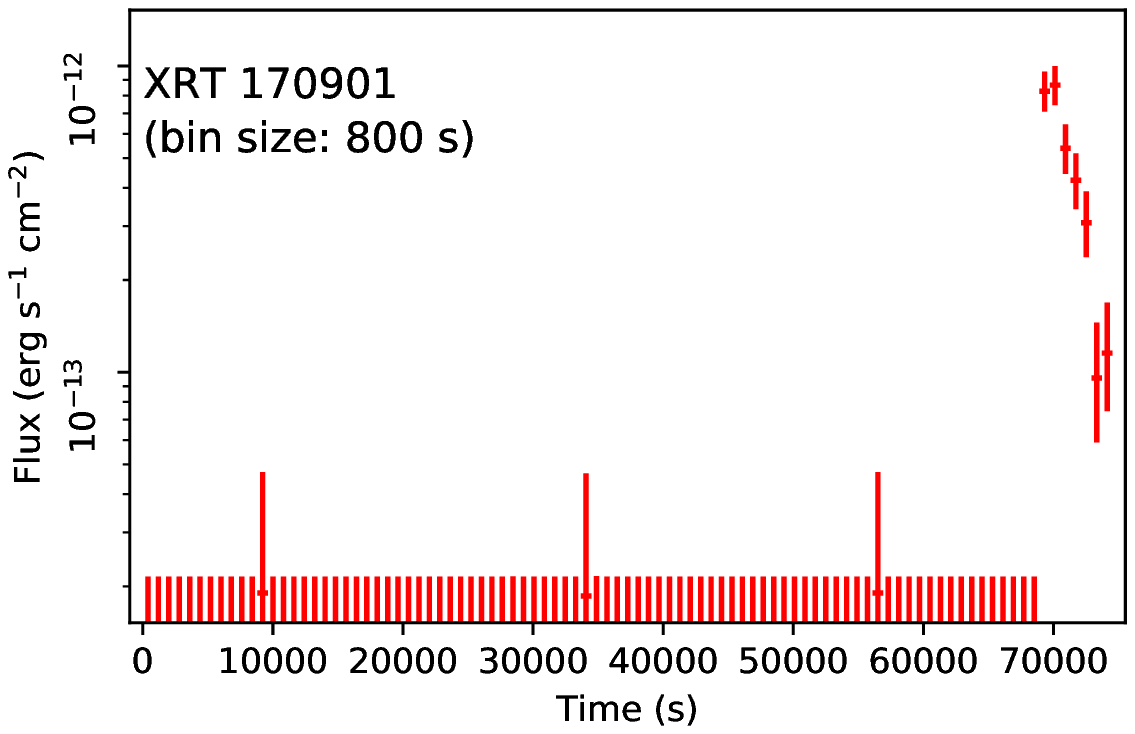}
\includegraphics[width=3.4in]{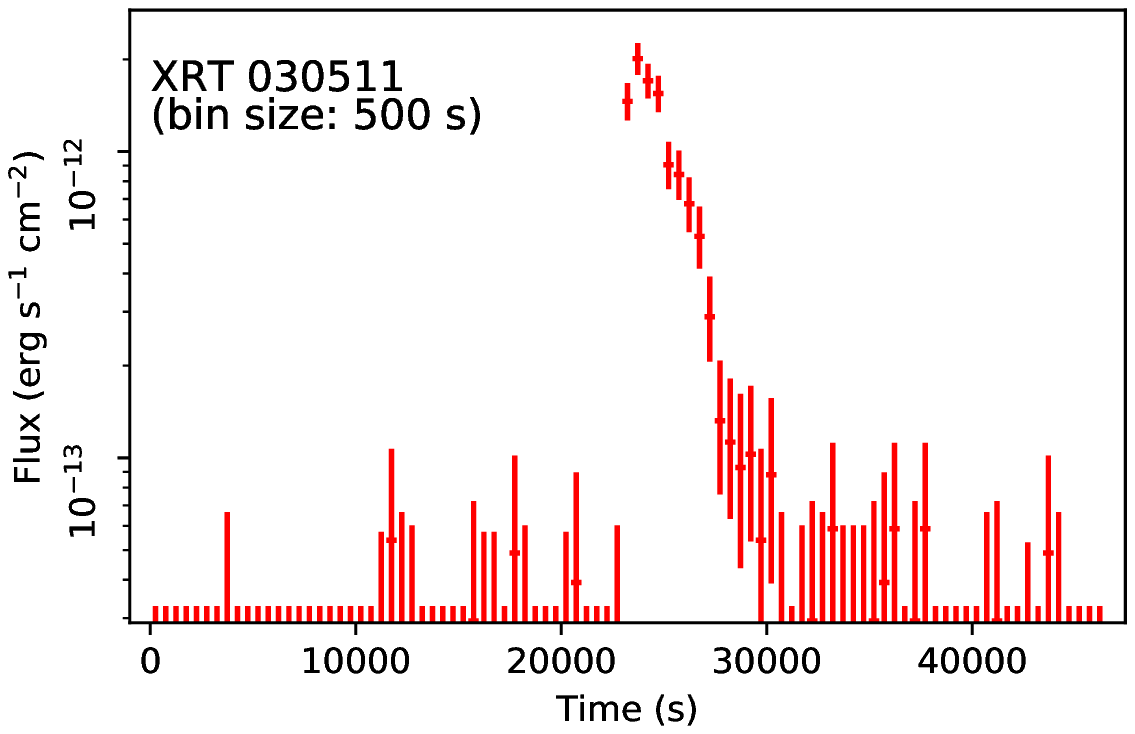}
\includegraphics[width=3.4in]{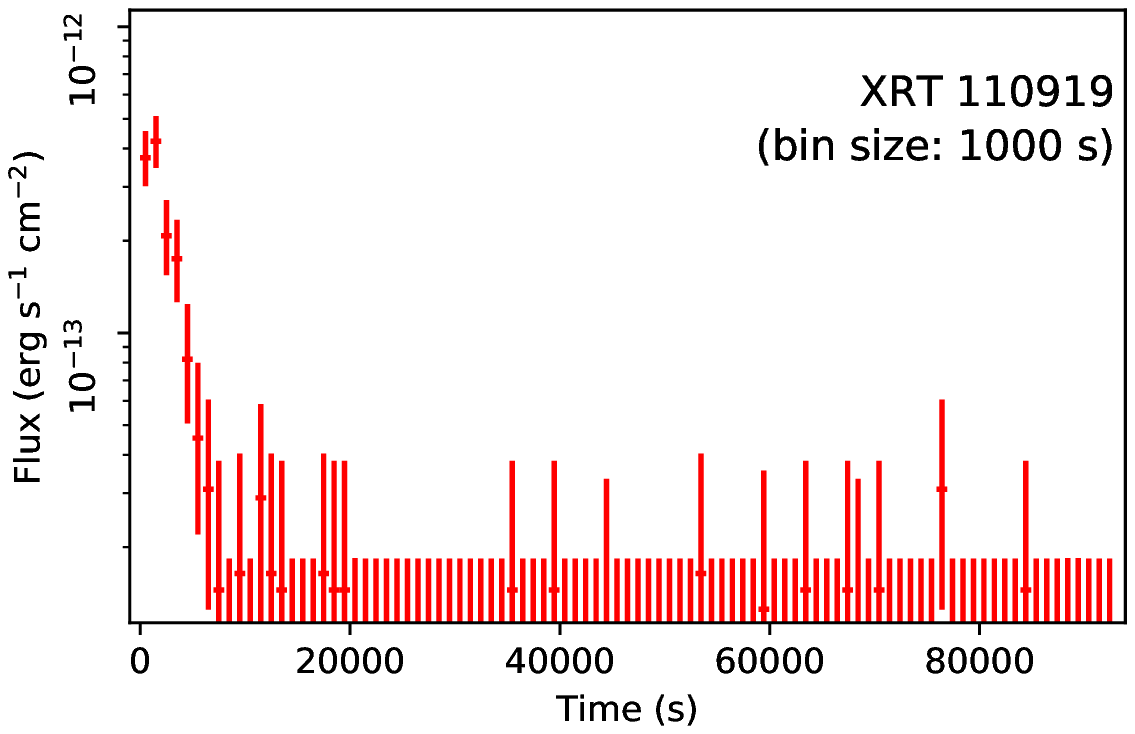}
\end{center}

\vskip -0.2in
\caption{The X-ray (0.5--7.0 keV) light curves of the three new FXTs
  from the \emph{Chandra} observations in which they are present. \label{fig:lc}}
\end{figure}

\begin{figure}
\begin{center}
  \includegraphics[trim=0 0.43in 0 0, clip,width=3.0in]{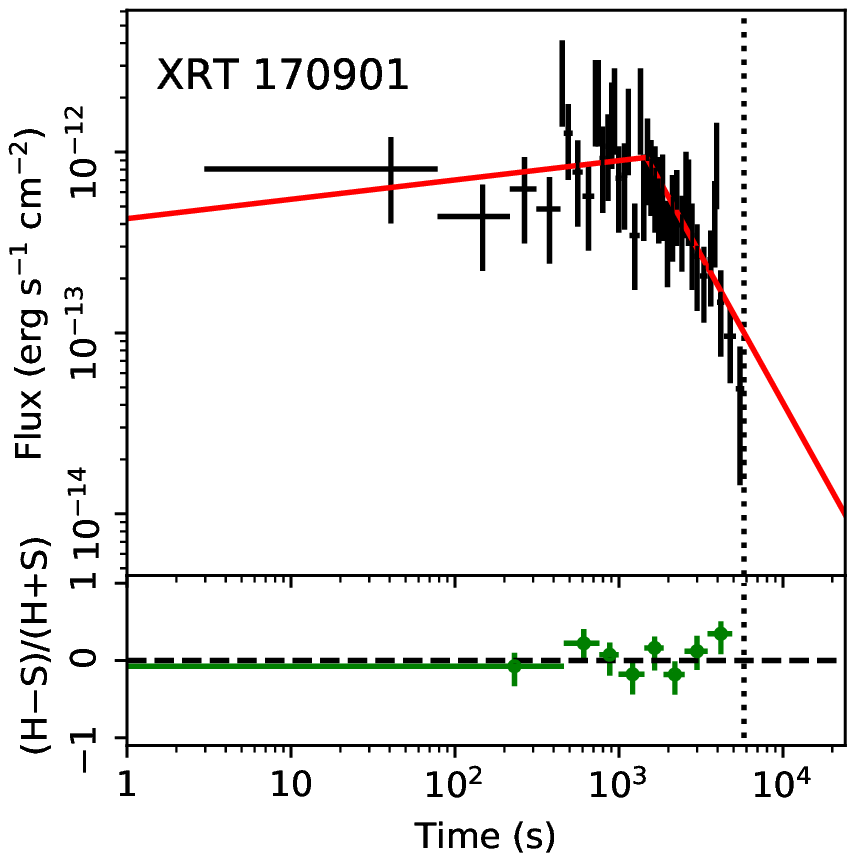}
\includegraphics[trim=0 0.43in 0 0, clip,width=3.0in]{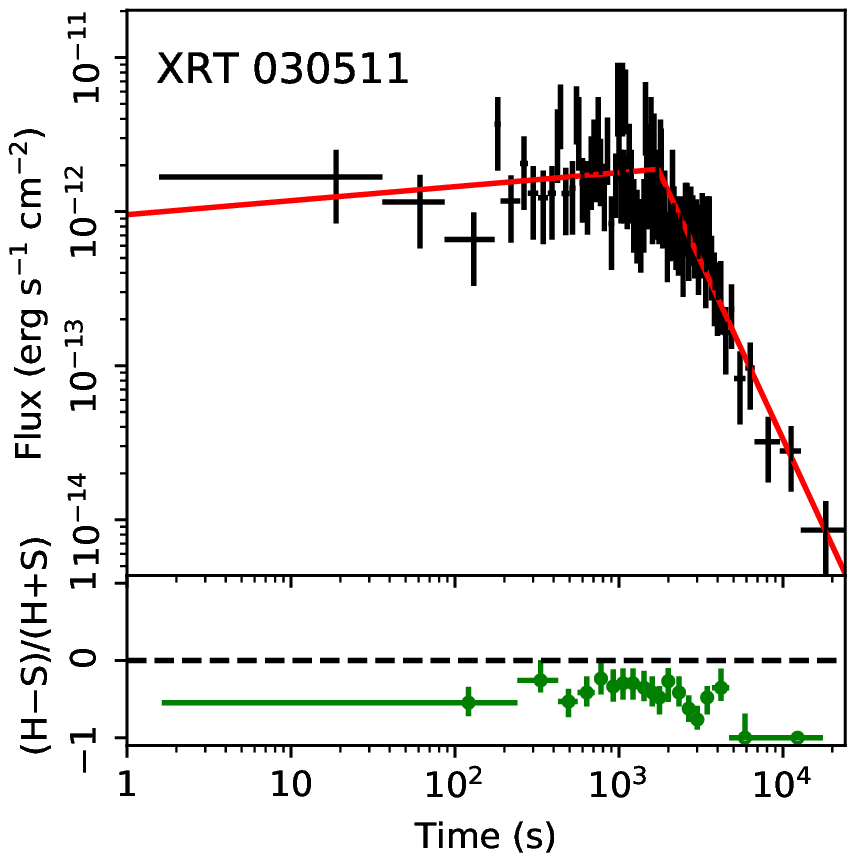}
\includegraphics[width=3.0in]{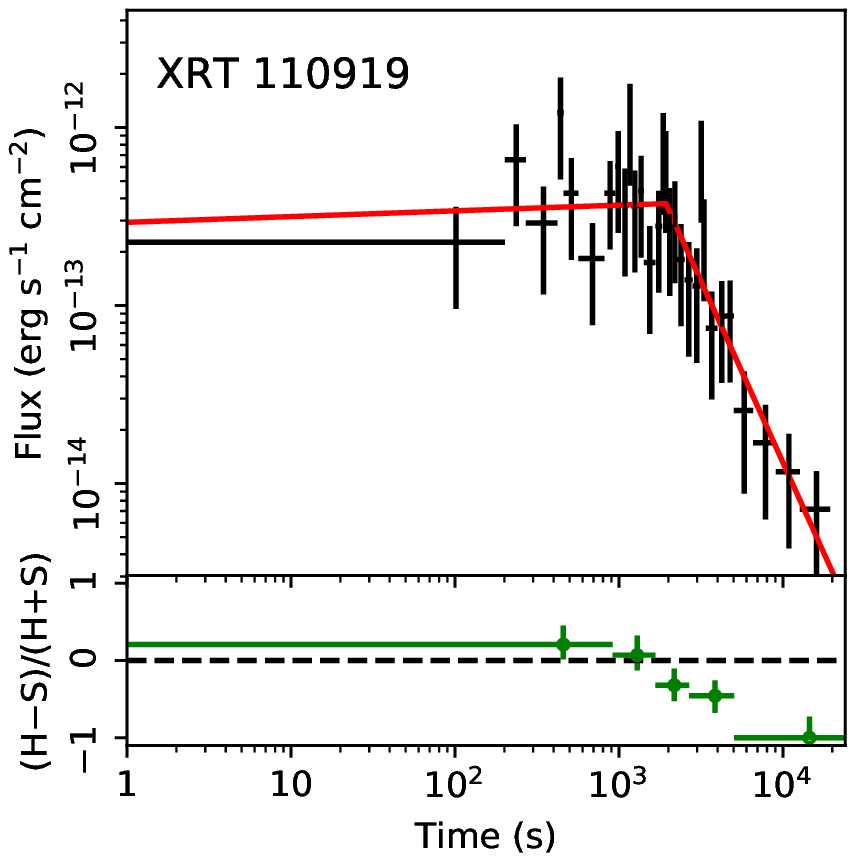}
\end{center}
\vskip -0.2in
\caption{The fits to the 0.5--7.0 keV light curves of the three new
  FXTs with a broken powerlaw (red solid lines) and the evolution of
  the hardness ratio over time. For
  viewing purposes, we have rebinned the light curves
  ($>2$$\sigma$ for XRT 170901 and XRT 030510 and $>$$1.5\sigma$
  for XRT 110919 in each bin). The hardness ratio was calculated with
  the requirement of at least 20 counts per time bin from the source region. \label{fig:lc2}}
\end{figure}

\begin{figure}
\begin{center}
  \includegraphics[trim=0 0.63in 0 0, clip,width=3.0in]{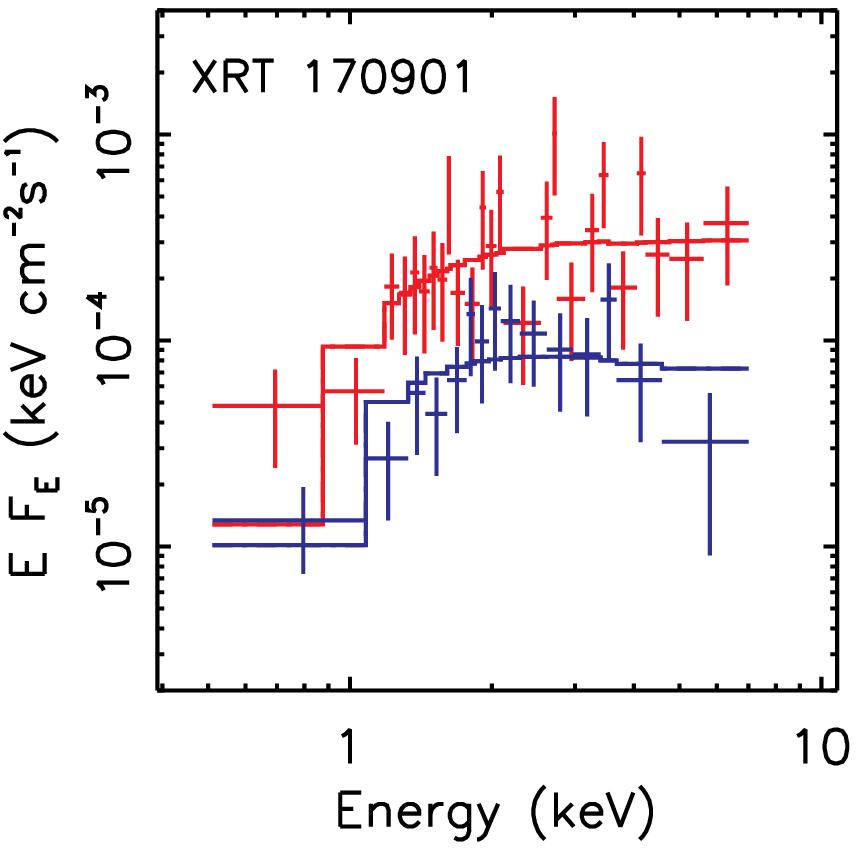}
\includegraphics[trim=0 0.63in 0 0, clip,width=3.0in]{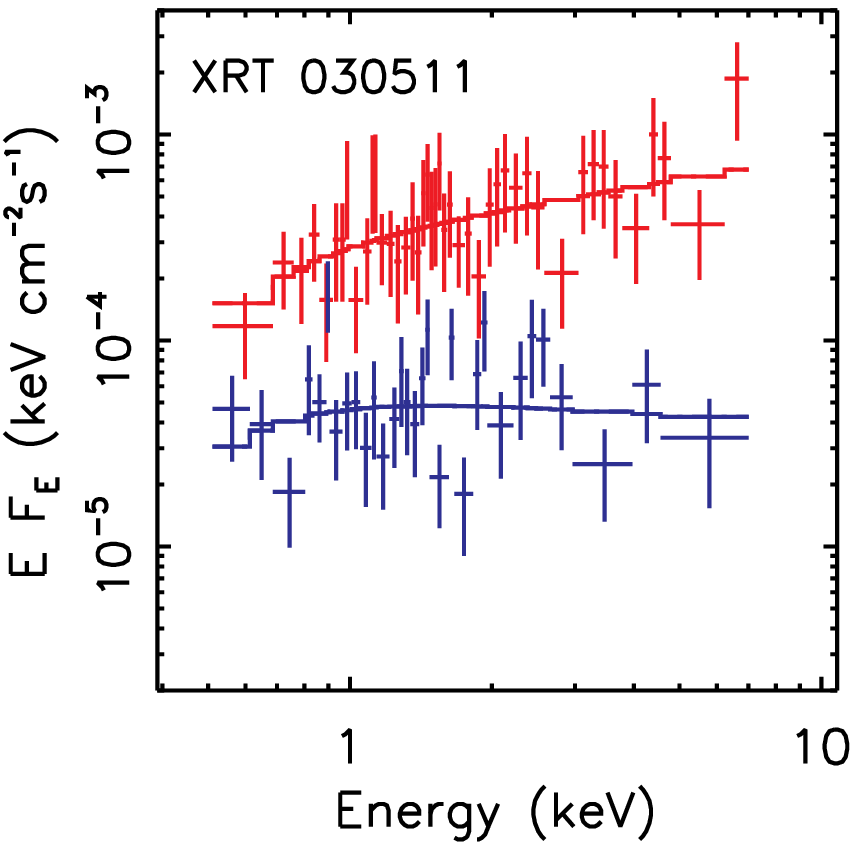}
\includegraphics[width=3.0in]{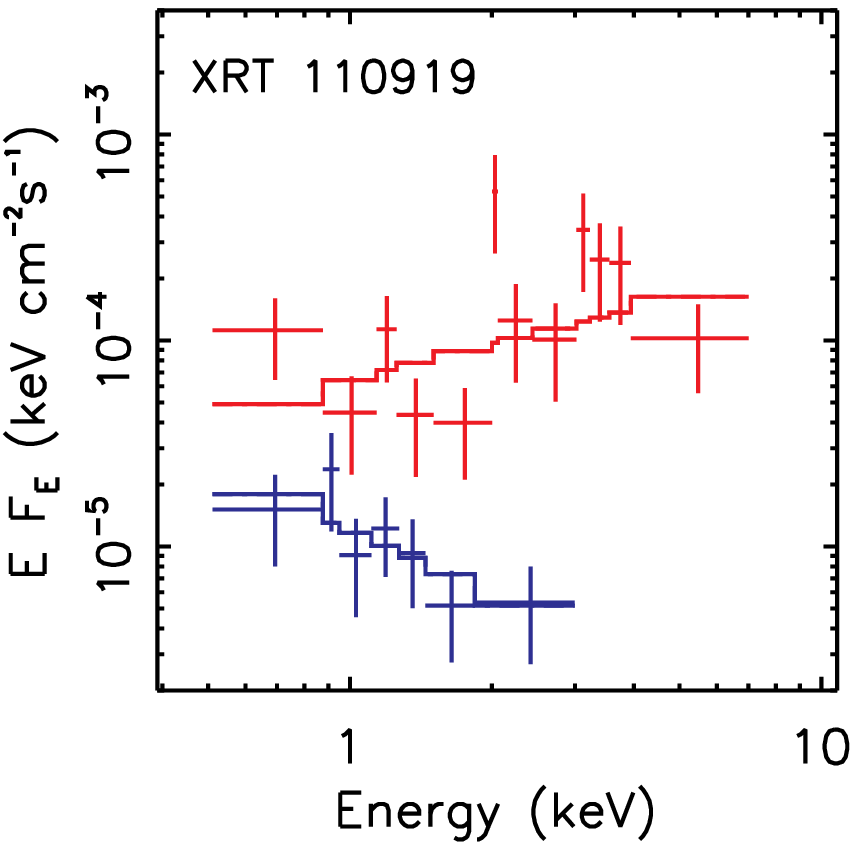}
\end{center}
\vskip -0.2in
\caption{The X-ray spectra of the three new FXTs in the plateau phase
  (red) and in
  the decay (blue), fitted with an absorbed powerlaw. \label{fig:spfit}}
\end{figure}

\begin{figure}
\begin{center}
  \includegraphics[width=3.0in]{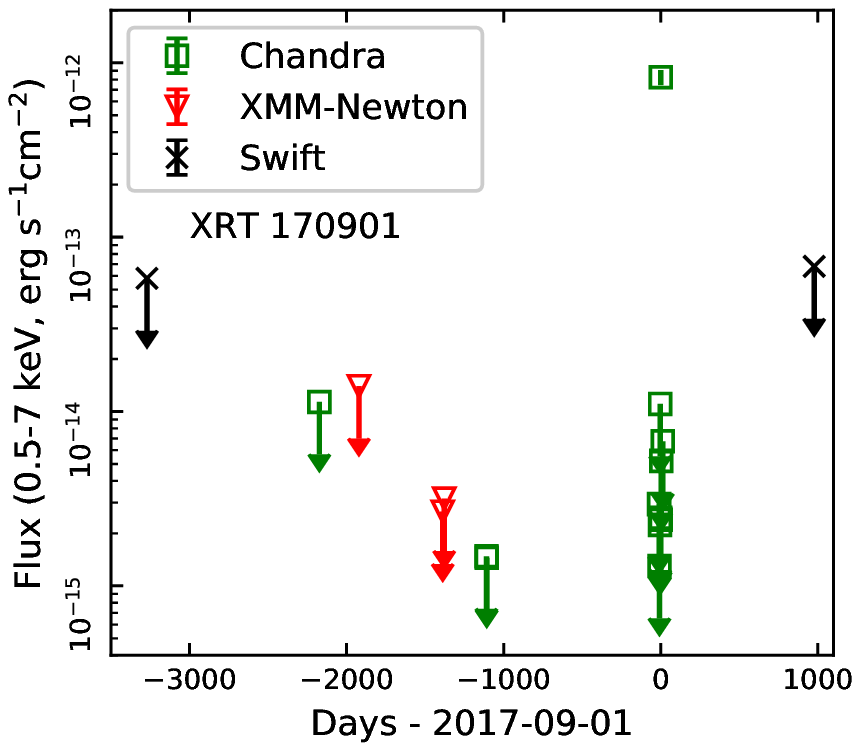}
\includegraphics[width=3.0in]{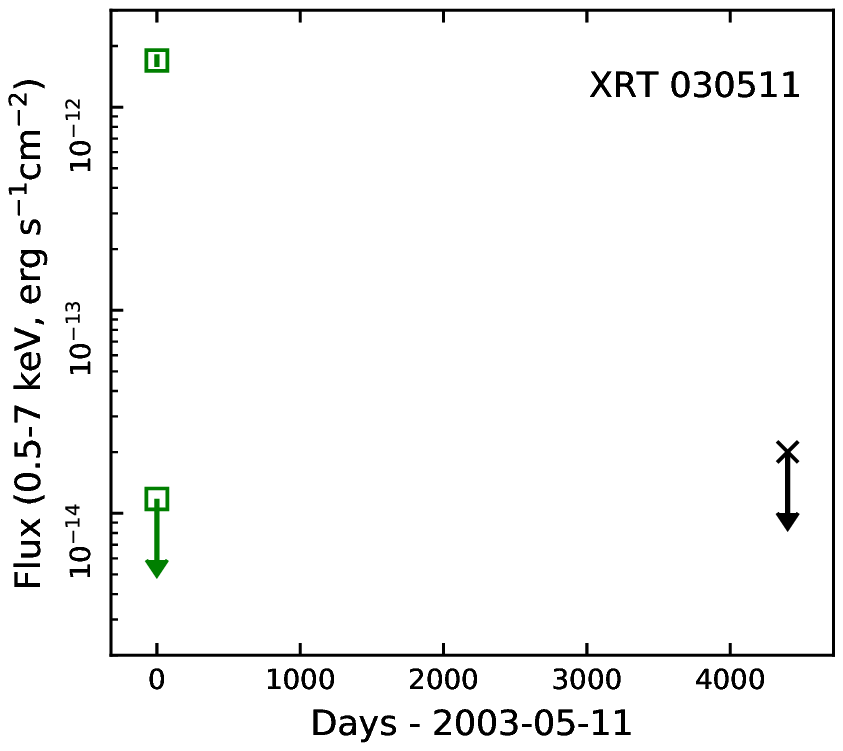}
\includegraphics[width=3.0in]{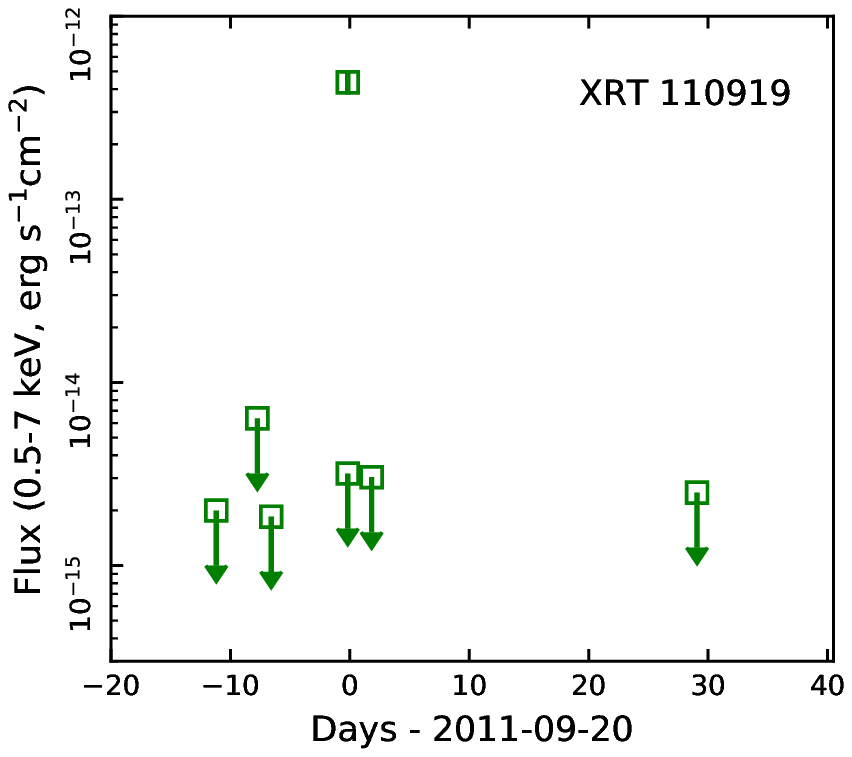}
\end{center}
\vskip -0.2in
\caption{The long-term 0.5--7.0 keV light curves of the three new FXTs
  from various X-ray observations. \label{fig:ltlc}}
\end{figure}

\begin{figure*}
\begin{center}
  \includegraphics[width=6.4in]{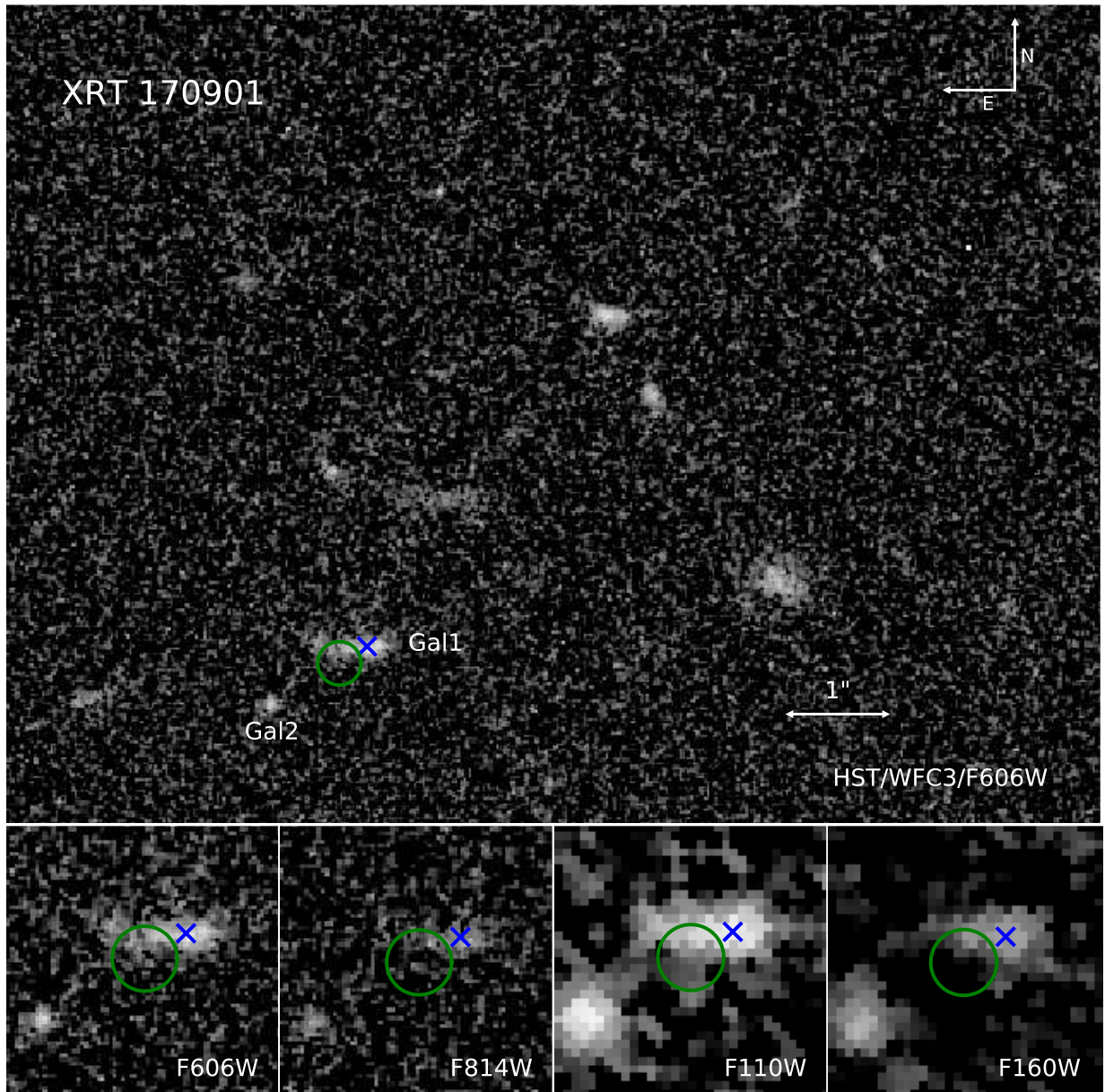}
\end{center}
\vskip -0.2in
\caption{The \emph{HST} F606W image 12\arcsec$\times$12\arcsec\ around XRT 170901,
  indicating the presence of a candidate highly elongated host
  galaxy (denoted as ``Gal1'' in the image). The lower panels zoom into this galaxy in various bands
  (2\arcsec$\times$2\arcsec\ in each panel). The 95\% X-ray positional
  uncertainty of this FXT is represented by a
  green circle of radius 0.3 arcsec in all panels. The blue cross
  corresponds to the peak pixel of F110W (R.A.=23:45:03.40,
  Decl.=$-$42:38:41.6). Another nearby galaxy (``Gal2'' in the
  figure, R.A.=23:45:03.50, Decl.=$-$42:38:42.2) is about 1 arcsec
  away from  XRT 170901, well outside the 95\% positional
  uncertainty of this FXT. \label{fig:hstxrt170901}}
\end{figure*}

\begin{figure*}
\begin{center}
  \includegraphics[width=6.4in]{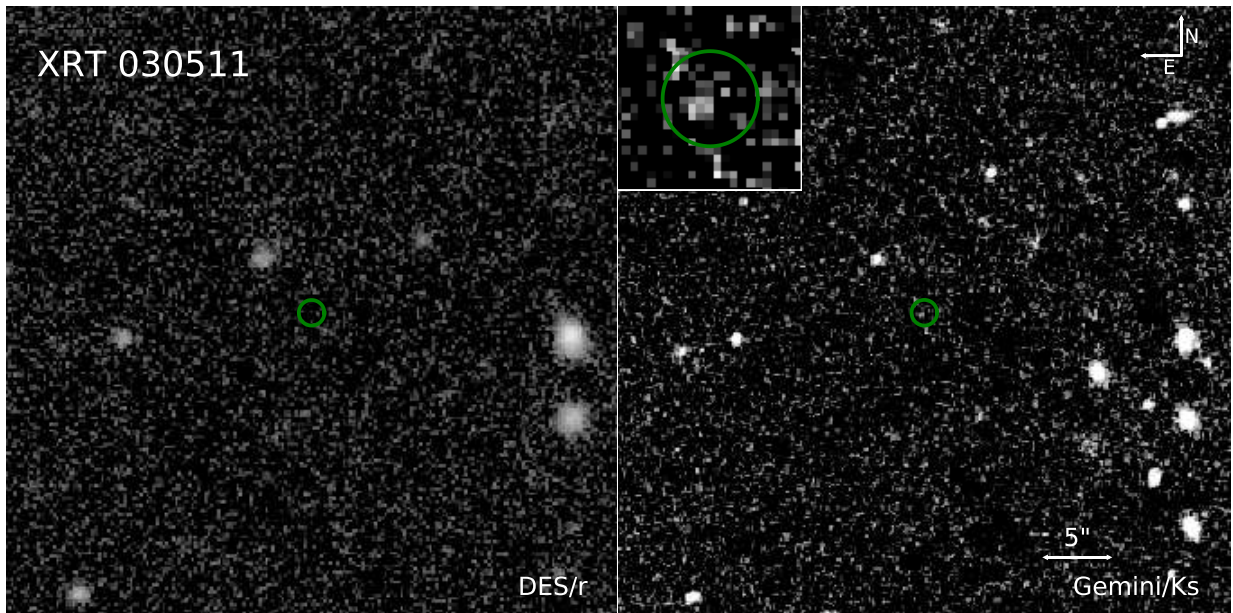}
\end{center}
\vskip -0.2in
\caption{The DES $r$-band and Gemini $K_\mathrm{s}$-band images around
  XRT 030511. The green circle of radius 1.05 arcsec represents its 95\% X-ray positional uncertainty. The inset in the
  right panel zooms into a 4\arcsec$\times$4\arcsec\ region around this FXT.
\label{fig:xrt030511optir}}
\end{figure*}

\begin{figure*}
\begin{center}
  \includegraphics[width=6.4in]{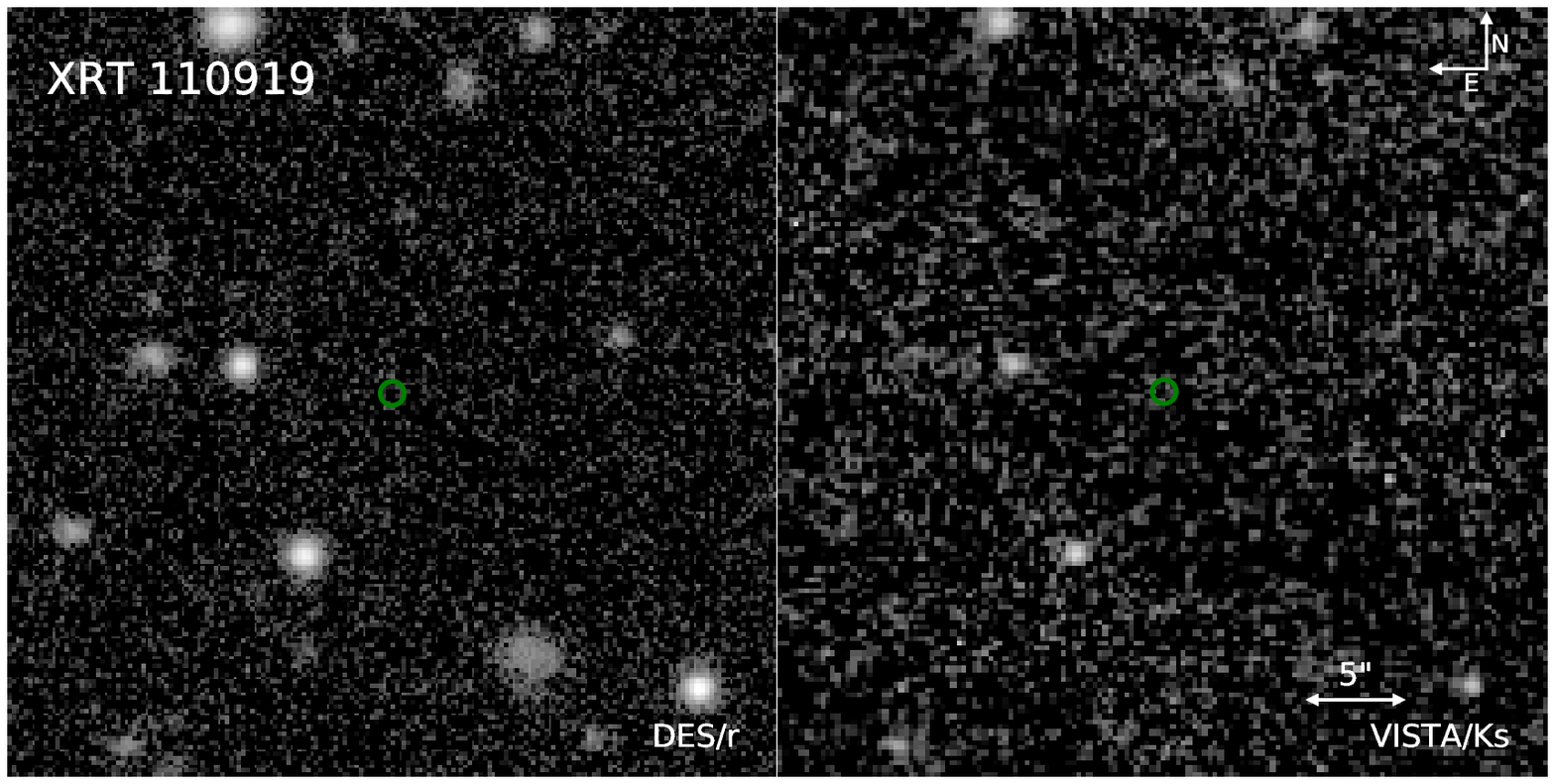}
\end{center}
\vskip -0.2in
\caption{The DES $r$-band and VISTA $K_\mathrm{s}$-band images around
  XRT 110919. The green circle of radius 0.78 arcsec represents its
  95\% X-ray positional uncertainty. 
\label{fig:xrt110919optir}}
\end{figure*}

\section{DATA ANALYSIS}
\label{sec:reduction}
\subsection{The Search Procedure}
Our search for FXTs is part of our large program to
  investigate all the X-ray sources serendipitously detected by
  \emph{Chandra} ACIS observations. In this program, we reduced each
  observation, carried out the source detection, performed
  multiwavelength cross-correlation, and classified the sources
  largely based on the method from \citet{liweba2012}. The details of
  this program will be reported in a future paper. In order to find
  the magnetar-powered FXT candidates, we selected out sources that
  are variable according to the CIAO tool glvary
  (variability index $\ge$ 6) from data up to 2019. Then we visually
  checked the light curves to identify the sources that showed a clear
  single flare in an observation. Most of these flaring sources are stars with low
  X-ray-to-IR flux ratios \citep{liweba2012} and can be ruled out
  securely. Only about a couple dozen were left at this step, and they
  included many interesting flaring sources that had been reported
  before
  \citep[][Xue19]{joglhe2013,gljofe2015, irmasi2016,batrsc2017}. We
  will focus on the three relatively bright FXTs that show similar
  properties (the flare has a fast rise and a peak plateau) to CDF-S
  XT2 and have not been studied yet.

\subsection{The X-ray Observations}
We analyzed not only the  \emph{Chandra} observations in which the
three FXTs were discovered but also other archival \emph{Chandra}, \emph{XMM-Newton},
and \emph{Swift} observations that happened to cover the fields
of these FXTs (Table~\ref{tbl:obslog}) in order to check whether
flares were also detected in other observations and, if not,
calculate the detection limit. We extracted the source and background light curves and
spectra and created the corresponding response files for spectral fits with the latest calibration (as of May 2019
for \emph{Swift} data, as of June 2020 for \emph{XMM-Newton} observations and CALDB
version 4.9.1 for \emph{Chandra} observations). Table~\ref{tbl:obslog} lists the size of
the source extraction region used for each observation; the background
spectra and light curves were extracted from a larger (by a factor of
6--90) background region near the
source. We used the software packages CIAO 4.12, SAS 19.0.0 and FTOOLS
6.25 for analysis. 

For the flares, we created energy spectra from the whole flare,
 from the peak, and from the decay separately, and fitted them with an absorbed
 single powerlaw using the XSPEC fitting package
\citep{ar1996}. Due to the low counts, we rebinned the spectra to have at
least one count in each bin (most bins have 1--3 counts after rebinning) and carried out the fits using the $C$
statistic. 

We also carried out the fits to the flare light curves with XSPEC. In
order to do this, we created the ``fake'' spectra
corresponding to the source and background light curves. Due to the
low counts, we also fitted the light curves with the $C$ statistic, with
the light curves binned to have at least one count in each bin.

In order to search for the multiwavelength counterparts to the FXTs,
we carried out the absolute astrometric correction to the X-ray
positions. We adopted the method in \citet{licawe2016}. The positions
of the X-ray sources were obtained with the CIAO tool
\verb|wavdetect|, with the statistical errors calculated following
\citet{kikiwi2007}. The relative astrometry of the X-ray sources is
corrected (about 0.5 arcsec for all cases) by matching with the IR sources from the VISTA Hemisphere
Survey \citep{mcbago2013}. In this step, we first identified candidate
matches and then searched for the translation and rotation of the
X-ray frame that minimize the total $\chi^2$ value ($\chi$ is the
ratio of the X-ray-IR separation to the total positional error) for
90\% of the candidate matches (we allowed 10\% of matches that have
the largest $\chi$ values to be
spurious or bad). The systematic positional errors from such an
astrometric correction procedure were estimated based on 200
simulations. All the positional errors of the X-ray sources that we
will report are the
root sum squares of the statistical and systematic errors.

\begin{deluxetable*}{lccc}
\tablecaption{Properties of the three new Chandra FXTs\label{tbl:properties}}
\tablehead{ & \colhead{CXO J234503.4$-$423841} &\colhead{CXO J050706.7$-$315210}&\colhead{CXO J010344.5$-$214845}\\
& \colhead{XRT 170901} & \colhead{XRT 030511} & \colhead{XRT 110919}}
\startdata

\multicolumn{4}{l}{X-ray position (J2000, astrometrically corrected):} \\
R.A., Decl.   & 23:45:03.44, $-$42:38:41.7 & 05:07:06.76, $-$31:52:10.8 & 01:03:44.59, $-$21:48:45.9\\
95\% error (arcsec) & 0.24 & 1.05 & 0.78\\
\hline
\multicolumn{4}{l}{Detected Chandra Observation:}\\
ObsID &20635 & 4062& 13454 \\
Observation Start Date & 2017-08-31 & 2003-05-10 & 2011-09-19 \\
Exposure (ks) & 74.2 & 46.2 & 91.8\\
Source region radius (arcsec) & 2.1 & 13.7 & 6.9 \\
Off-axis angle (arcmin) & 2.8 & 10.7 & 7.2 \\
$t_\mathrm{1st\, photon}$ (UTC) & 2017-09-01 13:26:47 & 2003-05-11 04:38:06 &2011-09-19 20:02:49\\
Flare net counts\tablenotemark{a} &164$\pm$13& 374$\pm$20 & 88.2$\pm$9.6\\
$T_{90}$ (ks)\tablenotemark{a} &$>$4.1$\pm$0.1&6.6$\pm$0.2&$<$11.0$\pm$0.2\\
\hline
\\
\multicolumn{4}{l}{Candidate host galaxy photometry\tablenotemark{a}:} \\
\hline
Optical (AB mag)& F606W=$24.92\pm0.01$& $g>25.6$, $r>24.7$, $i>24.2$&$g>25.2$, $r>24.6$, $i>24.2$ \\
&F814W=$25.00\pm0.02$&$z>23.9$, $Y>22.8$&$z>23.9$, $Y>22.7$\\
IR (AB mag)& F110W=$24.81\pm0.01$&$J>21.7$ & $J>22.1$ \\
& F160W=$24.64\pm0.01$&$K_\mathrm{s}=24.5\pm0.2$&$K_\mathrm{s}>20.1$\\
\hline
\\
Light curve fits\tablenotemark{a}: \\
\hline
\multicolumn{4}{l}{Broken powerlaw:}\\
Initial index &$0.11^{+0.33}_{-0.13}$&$0.09\pm0.08$& $0.03\pm0.15$\\
Break (ks) &$1.53^{+0.31}_{-0.59}$&$1.73\pm0.13$& $1.93^{+0.25}_{-0.31}$\\
Second index &$-1.63^{+0.53}_{-0.36}$&$-2.30\pm0.16$&$-2.04^{+0.25}_{-0.30}$ \\
\hline
\multicolumn{4}{l}{Magnetar model (free $n$):}\\
$\tau$ (ks) & $>24.0$ & $>35.3$ & $11.6^{+46.6}_{-6.4}$\\
$n$ & $<1.4$ & $<1.2$& $1.7\pm0.5$ \\
C-stat (dof)  & 155.6 (161) & 319.8 (346) & 93.6 (88) \\
\hline
\multicolumn{4}{l}{Magnetar model ($n=3$, fixed):}\\
$\tau$ (ks) & $3.0\pm0.6$ & $1.8\pm0.2$ & $2.2\pm0.4$\\
C-stat (dof)  & 162.6 (162) & 379.0 (347) & 99.1 (89) \\
\hline
\multicolumn{4}{l}{Magnetar model ($n=5$, fixed):}\\
$\tau$ (ks) & $1.0\pm0.3$ & $0.3\pm0.1$ & $0.4\pm0.1$ \\
C-stat (dof) &  170.1 (162) & 508.6 (347) & 117.7 (89)\\
\hline
\\
\multicolumn{4}{l}{Fits to the whole flare
  spectrum\tablenotemark{b}:}\\
\hline
Exposure (ks) & 5.7 & 13.6 & 19.5 \\
\hline
\multicolumn{4}{l}{Single powerlaw ($z=0$)}\\
$N_\mathrm{H,i}$ (10$^{21}$ cm$^{-2}$) & $8.8^{+3.7}_{-3.3}$ & $0.8^{+1.3}_{-0.8}$ & $0.0^{+0.6}$\\
$\Gamma$ & $2.16\pm0.31$ & $1.84\pm0.27$ & $1.88\pm0.21$\\
\hline
\multicolumn{4}{l}{Single powerlaw ($z=1.0$)}\\
$N_\mathrm{H,i}$ (10$^{21}$ cm$^{-2}$) & $46.4^{+22.1}_{-19.4}$ & $3.6^{+6.3}_{-3.6}$ & $0.0^{+2.8}$\\
$\Gamma$ & $2.12\pm0.31$ & $1.81\pm0.25$ & $1.88\pm0.21$\\
\hline
\\
\multicolumn{4}{l}{Joint fits to the plateau and decay spectra:}\\
\hline
Plateau/decay exposures (ks) & 2.0/3.7 & 1.7/11.9  & 2.2/17.3 \\
Plateau/decay net counts & 104/61 & 203/167 & 52/37\\
\hline
\multicolumn{4}{l}{Single powerlaw ($z=0$)}\\
$N_\mathrm{H,i}$  (10$^{21}$ cm$^{-2}$) & $7.6^{+3.7}_{-3.2}$ & $1.0^{+1.3}_{-1.0}$ & $0.0^{+0.9}$\\
$\Gamma$ Plateau/decay & $2.05\pm0.20$/$2.29\pm0.27$ & $1.65\pm0.21$/$2.17\pm0.27$&$1.45\pm0.27$/$3.05\pm0.42$ \\
\hline
\multicolumn{4}{l}{Single powerlaw ($z=1.0$)}\\
$N_\mathrm{H,i}$  (10$^{21}$ cm$^{-2}$) & $43.9^{+21.9}_{-19.3}$ & $4.2^{+6.5}_{-4.1}$ & $0.0^{+3.5}$\\
$\Gamma$ Plateau/decay & $2.03\pm0.21$/$2.27\pm0.27$ &
$1.61\pm0.21$/$2.13\pm0.27$& $1.45\pm0.27$/$3.05\pm0.42$ \\
\hline
\\
\multicolumn{4}{l}{High-energy emission upper limit:} \\
\hline
sGRB Fluence ($10^{-7}$ erg cm$^{-2}$) & 5.8 & 5.1 & 5.3 \\
lGRB peak flux ($10^{-7}$ erg cm$^{-2}$ s$^{-1}$)  & 1.6 & 1.27 & 1.63 
\enddata 
\tablecomments{\textsuperscript{a}Errors or limits are at the $1\sigma$ confidence
  level, except for the magnitude limits, which refer to the magnitude of the 90th faint source
  among the 100 nearest sources. \textsuperscript{b}Errors are at the
  90\% confidence level. }
\end{deluxetable*}

\tabletypesize{\scriptsize}
\setlength{\tabcolsep}{0.02in}
\begin{deluxetable*}{rccccccc}
\tablecaption{The X-ray Observation Log\label{tbl:obslog}}
\tablewidth{0pt}
\tablehead{\colhead{Obs. ID} &\colhead{Date} & \colhead{Detector}
  &\colhead{OAA} &\colhead{Expo} &\colhead{$r_\mathrm{src}$}  &
   \colhead{0.5--7 keV count rate} & \colhead{0.5--7 keV flux} \\
 & & & & (ks)& & ($10^{-3}$ counts s$^{-1}$) & (erg s$^{-1}$ cm$^{-2}$)\\
 (1) & (2) &(3) & (4) & (5) & (6) & (7) & (8) 
}
\startdata
\multicolumn{4}{l}{XRT 170901}\\
\hline
\multicolumn{4}{l}{\emph{Swift}:}\\
00038092001&2008-09-16&XRT&10.9$\arcmin$&2.7&20$\arcsec$&\multirow{6}{*}{$<$0.99} &\multirow{6}{*}{$<$5.8e-14}\\
00038092002&2008-09-17&XRT&10.9$\arcmin$&4.5&20$\arcsec$&\\
00038092004&2008-11-06&XRT&11.2$\arcmin$&0.1&20$\arcsec$&\\
00038092005&2008-11-09&XRT&9.8$\arcmin$&0.6&20$\arcsec$&\\
00038092006&2009-03-29&XRT&7.2$\arcmin$&0.3&20$\arcsec$&\\
00038092007&2009-03-31&XRT&6.1$\arcmin$&1.2&20$\arcsec$&\\
\hline
00081309001&2020-05-04&XRT&7.9$\arcmin$&1.5&20$\arcsec$&\multirow{2}{*}{$<$1.32} &\multirow{2}{*}{$<$6.8e-14}\\
00081309002&2020-05-05&XRT&6.9$\arcmin$&3.0&20$\arcsec$&\\ 
\multicolumn{4}{l}{\xmm:}\\
0693661801 & 2012-05-28 & pn & 5.9$\arcmin$ & 8.9 & 15$\arcsec$ &$<$2.4 &$<$1.4e-14\\
0722700101 & 2013-11-22 & pn & 5.8$\arcmin$ & 104.4 & 15$\arcsec$ &$<$0.46 &$<$3.2e-15\\
0722700201 & 2013-11-13 & pn & 5.8$\arcmin$ & 81.5 & 15$\arcsec$ &$<$0.41 &$<$2.7e-15\\
\multicolumn{4}{l}{\textit{Chandra}:}\\
13401&2011-09-19&ACIS-I&4.4$\arcmin$& 11.9 & 2$\farcs$5 &$<$0.66&$<$1.1e-14\\
16135&2014-08-18&ACIS-I&2.5$\arcmin$& 58.1 & 1$\farcs$2 &$<$0.10&$<$1.5e-15\\
16545&2014-08-20&ACIS-I&2.5$\arcmin$& 59.2 & 1$\farcs$2 &$<$0.10&$<$1.4e-15\\
19581&2017-08-23&ACIS-I&2.4$\arcmin$& 88.8 & 1$\farcs$1 &$<$0.07&$<$1.3e-15\\
19582&2017-08-28&ACIS-I&2.6$\arcmin$& 10.0 & 1$\farcs$2 &$<$0.59&$<$1.1e-14\\
19583&2017-09-14&ACIS-I&3.6$\arcmin$& 24.7 & 1$\farcs$9 &$<$0.31&$<$6.7e-15\\
20630&2017-08-21&ACIS-I&2.3$\arcmin$& 38.5 & 1$\farcs$1 &$<$0.15&$<$2.9e-15\\
20631&2017-08-27&ACIS-I&2.4$\arcmin$& 50.4 & 1$\farcs$1 &$<$0.12&$<$2.2e-15\\
20634&2017-08-29&ACIS-I&2.6$\arcmin$& 70.6 & 1$\farcs$2 &$<$0.14&$<$2.5e-15\\
20635&2017-08-31&ACIS-I&2.7$\arcmin$& 68.5 & 2$\farcs$1 &$<$0.16&$<$2.4e-15\\
20636&2017-09-04&ACIS-I&2.8$\arcmin$& 21.3 & 1$\farcs$3&$<$0.28&$<$5.2e-15\\
\hline
\multicolumn{4}{l}{XRT 030511}\\
\hline
\multicolumn{4}{l}{\emph{Swift}:}\\
00084253001&2015-04-27&XRT&11.7$\arcmin$&0.7&20$\arcsec$&\multirow{17}{*}{$<$0.36} &\multirow{17}{*}{$<$2.0e-14}\\
00084253002&2015-05-18&XRT&10.3$\arcmin$&2.0&20$\arcsec$&\\
00084253003&2015-05-19&XRT&10.2$\arcmin$&1.3&20$\arcsec$&\\
00084253004&2015-05-23&XRT&10.8$\arcmin$&0.7&20$\arcsec$&\\
00084253005&2015-05-25&XRT&7.8$\arcmin$&3.4&20$\arcsec$&\\
00084253006&2015-05-27&XRT&7.4$\arcmin$&0.6&20$\arcsec$&\\
00084253007&2015-05-28&XRT&7.8$\arcmin$&0.5&20$\arcsec$&\\
00084253008&2015-06-02&XRT&9.4$\arcmin$&0.5&20$\arcsec$&\\
00084253010&2015-06-07&XRT&9.0$\arcmin$&2.9&20$\arcsec$&\\
00084254001&2015-05-06&XRT&11.6$\arcmin$&0.6&20$\arcsec$&\\
00084254002&2015-05-25&XRT&8.7$\arcmin$&0.2&20$\arcsec$&\\
00084254003&2015-06-07&XRT&9.0$\arcmin$&5.0&20$\arcsec$&\\
00084254004&2015-06-09&XRT&8.5$\arcmin$&4.8&20$\arcsec$&\\
00084254005&2015-06-12&XRT&9.5$\arcmin$&1.2&20$\arcsec$&\\
00084254006&2015-06-17&XRT&11.4$\arcmin$&1.4&20$\arcsec$&\\
00084254007&2015-06-19&XRT&10.8$\arcmin$&0.9&20$\arcsec$&\\
00084254008&2015-06-21&XRT&11.3$\arcmin$&0.9&20$\arcsec$&\\
\multicolumn{4}{l}{\textit{Chandra}:}\\
4062&2003-05-10&ACIS-S&10.7$\arcmin$& 22.9 &13$\farcs$7&$<$0.72&$<$1.2e-14\\
\hline
\multicolumn{4}{l}{XRT 110919}\\
\hline
\multicolumn{4}{l}{\textit{Chandra}:}\\
13447&2011-09-08&ACIS-I&5.2$\arcmin$& 69.1 & 3$\farcs$2 &$<$0.14&$<$2.0e-15\\
13448&2011-09-13&ACIS-I&7.2$\arcmin$& 146.0 & 5$\farcs$7 &$<$0.06&$<$1.8e-15\\
13454&2011-09-19&ACIS-I&7.2$\arcmin$& 72.2 & 6$\farcs$9&$<$0.19&$<$3.2e-15\\
13455&2011-10-19&ACIS-I&6.8$\arcmin$& 69.6 & 5$\farcs$3 &$<$0.16&$<$2.5e-15\\
14343&2011-09-12&ACIS-I&7.2$\arcmin$& 35.3 & 5$\farcs$7 &$<$0.24&$<$6.4e-15\\
14346&2011-09-21&ACIS-I&7.2$\arcmin$& 85.3 & 5$\farcs$7 &$<$0.16&$<$3.0e-15
\enddata 
\tablecomments{Columns: (1) the observation ID, (2) the observation
  start date, (3) the detector, (4) the off-axis angle, (5) the clean
  exposure time of the data used in the final analysis after excluding
  periods of
  strong background flares, (6) the radius of the
  source circular extraction region, (7) the $3\sigma$ upper limit net
  count rate, (8) the $3\sigma$ upper limit of the 0.5--7 keV
  flux. The limits are either from individual \emph{Chandra} and \emph{XMM-Newton} observations
    or from combinations of
    \emph{Swift} observations.  The count rate upper limits were calculated with the CIAO task
  \textit{aprates}, which uses the Bayesian approach.  The
flux upper limits were calculated, assuming an absorbed powerlaw
spectrum inferred from the fit to the whole flare
spectrum.  For the three \emph{Chandra} observations 20635,
  4062, and 13454 in which the FXTs detected, the exposures, the count
  rate and flux upper limits refer to the persistent periods.}
\end{deluxetable*}

\subsection{Optical and IR images}
The field of XRT 170901 happened to be covered in the \emph{Hubble Space Telescope} (\emph{HST}) WFC3 Infrared
Spectroscopic Parallel Survey \citep{atmamc2010}, with two optical
images (F606W and F814W, 2600 s each) taken on 2014 July 11 and two IR images
(F110W and F160W, 1518 s and 759 s, respectively) taken on 2014 July
10. We produced the drizzled
images with the DrizzlePac software package. The pixel size was set to
be 0.03 arcsec for the optical images and 0.06 arcsec for the
IR images. The astrometry of these \emph{HST} images was aligned
to that of the VISTA Hemisphere Survey by cross-matching the point sources detected in
both frames.

We also obtained a deep Gemini Flamingos-2 $K_\mathrm{s}$ image (Program ID
GS-2019B-FT-208) around the field of XRT 030511 in
order to search for the counterpart. The IR filter was adopted because
the counterpart might be very red if the FXT is associated with a
high-redshift passive galaxy or with
a very late-type star. Due to weather constraints, individual exposures ranging
from 12 s to 20 s were taken over several nights spanning about a
month (2019-11-30 to 2020-01-01) under the best
seeing conditions (average seeing 0.45 arcsec) at the Gemini South observatory. We reduced the images
using the Gemini DRAGONS software. The final stacked image obtained
has a total exposure of 1926 s. The astrometry of the Gemini image was also
aligned with the VISTA survey.

\section{RESULTS}
\label{sec:res}

\subsection{The X-ray Flares}
Table~\ref{tbl:properties} lists the various properties of the three
FXTs that we discovered. 
Figure~\ref{fig:lc} plots the background-subtracted light curves of the three FXTs
from the \emph{Chandra} observations in which they were detected. XRT
170901  was detected at 69 ks into the observation, and the observation was
stopped before the flare ended. XRT 030511 was detected at the middle
of the observation and seemed to end before the observation was
stopped. XRT 110919 occurred at the beginning of the observation and
probably began earlier than the observation. We calculated $T_{90}$, the timespan from the
5\%-th to 95\%-th of the total detected counts and obtained $4.1$ ks,
6.6 ks and 11.0 ks for XRT 170901, XRT 030511, and XRT 110919,
respectively (Table~\ref{tbl:properties}). The value for XRT 170901 should be a lower limit because
we missed the long faint decay for this FXT, and the value for XRT
110919 is most likely an upper limit because we probably missed the
early bright phase of this FXT.  No clear rising phase was detected in 
XRT 170901 and XRT 030511. We estimated the upper limit of
the rise time for these two FXTs following Xue19, which assumed a linear
rise profile and the non-detection of the rising period. We obtained a
$1\sigma$ upper limit on the rise time of 42 s and 19 s for XRT 170901
and XRT 030511, respectively.

Figure~\ref{fig:lc2} zooms in on the flares. The time axis is on a
logarithmic scale with the time zero set at the start of the flares.
The start times of the flares were estimated as follows. We first
identified the first photon detected from each flare, and then
calculated the average waiting time $\mathrm{d}t$ of the next nine
photons from the source region. We assumed the flare start time to be
at $\mathrm{d}t/2$ before the first detection. We estimated the chance
probability $p_\mathrm{bg}$ that the first flare photon that we
identified is in fact from the background as
$R_\mathrm{bg}\mathrm{d}t$, where $R_\mathrm{bg}$ is the background
count rate in the source region. We obtained
$p_\mathrm{bg}\sim0.03$\%, 1.6\%, and 0.8\% for XRT 170901, XRT
030511, and XRT 110919, respectively. In comparison, the time
  gap between the assumed first flare photon and the photon
immediately before it (also from the source region) for each case is large,
12 ks and 430 s, with $p_\mathrm{bg}$
$\sim15$\% and 35\%, for XRT 170901 and XRT 030511, respectively. They
are most likely from the background.  The first flare photon that we
identified for XRT 110919 is the first photon of the observation
in the source region. Therefore, our identification of the first flare
photon is fairely secure.

We first fitted the light curves with a broken powerlaw as was done in
Xue19. The results are given in
Table~\ref{tbl:properties} and shown in
Figure~\ref{fig:lc2}. The fits started from the beginning of
  the flares until the end of the observation (for XRT 170901) or the end of the flares (for XRT 030511 and XRT 110919, at around 20 ks).  The initial index was inferred to be
consistent with zero in all three XRTs. The breaks are all between
1.5--2.0 ks. The second indexes were all consistent with 2.0. We note
that for XRT 030511, which occurred at the beginning of the
observation, the break could occur at a later time than that we inferred
in Table~\ref{tbl:properties}, which assumed the flare started at the
beginning of the observation.

We also fitted the light curves with the magnetar model
$F_{\rm X}(t)\propto (1+t/\tau)^{4/(1-n)}$, in which $n$ is the
braking index and $\tau$ is the spindown timescale of the
magnetar. This model was commonly used to fit the X-ray afterglow of
GRBs \citep[e.g.,][]{lulali2019}. The results are given in
Table~\ref{tbl:properties}.  For XRT 110919, we inferred $n=1.7\pm0.5$
and $\tau=11.6^{+46.6}_{-6.4}$ ks. However, for both XRT 170901 and
XRT 030511, we obtained $n$ very close to 1.0 and $\tau$ very large,
which is caused by the degeneracy of these two parameters when $n$ is
very close to 1.0. In this case, the model becomes an exponential
function $\exp(-t/\tau)$. We are most concerned about whether good
fits can be obtained with the standard magnetar models of $n=3$ (the
magnetic dipole braking dominated) or $n=5$ (GW braking
dominated). Therefore, we also fitted the models with $n=3$ and 5 and
compared with the fits with free $n$. The results are given in
Table~\ref{tbl:properties}. The fits with $n=5$ are strongly
disfavored, with the $C$ statistic values larger than those of the
fits of free $n$ by 14.5, 188.8, and 24.1 for XRT 170901, XRT 030511,
and XRT 110919, respectively. The fits with $n=3$ are much better,
with the $C$ statistic values larger than those of the fits of free
$n$ by 7.0, 59.2, and 5.5, respectively, but at least for the case of
XRT 030511, the improvement is still not enough. Therefore all the
three FXTs might prefer braking indexes smaller than $3$, as was found
in CDF-S XT2 \citep{xizhda2019} and many GRBs \citep{stdada2018}. This
could imply the presence of other braking mechanisms such as fall-back
accretion onto the magnetar \citep{mebegi2018}.

  The fits to the whole flare spectra with an absorbed single
  powerlaw, assuming two cases of redshifts $z=0.0$ and $z=1.0$,
  inferred a photon index of $\sim$2.0 in all the three FXTs (Table~\ref{tbl:properties}). While no significant absorption was
  inferred in XRT
  030511 or XRT 110919,  XRT 170901 seems to be heavily absorbed
 (intrinsic absorption column density $N_\mathrm{H,i}=0.9\pm0.4\times10^{22}$ cm$^{-2}$ and
 $4.6_{-1.9}^{+2.2}\times10^{22}$ cm$^{-2}$ when $z=0.0$ and 1.0 was
 assumed, respectively). While the photon index
  inferred is fairly independent of the redshift assumed, the
  absorption column density inferred in XRT 170901 increased with the redshift
  assumed.

  Figure~\ref{fig:lc2} shows the temporal evolution of the hardness
  ratio for all three FXTs. We followed Xue19 to define the hardness
  ratio as $(H-S)/(H+S)$, where $S$ and $H$ are the count rates from
  the soft energy band 0.5--2.0 keV and the hard energy band
  2.0--7.0 keV, respectively. In order to have enough statistics, the hardness ratios were calculated with
  varying time bin size, requiring that each data point has at least 20 counts
  from the source region. The uncertainties of the hardness ratios
  were calculated with the Bayesian code \verb|BEHR|
  \citep{pakasi2006}, due to the low statistics. We found that the
  hardness ratios in both XRT 030511 and XRT 110919 decreased
  significantly, indicating spectral softening, in the decay in
  these two FXTs. The hardness ratios in XRT 170901 seem to remain
  constant over time when the flare was observed.

  In order to characterize the spectral variation
  further, we fitted the plateau and decay spectra simultaneously with $N_\mathrm{H,i}$ tied to be the same for each
  FXT. The fitting results are shown in Figure~\ref{fig:spfit} and are given in Table~\ref{tbl:properties}. We
  note that the uncertainties on the photon indexes in the table were
  obtained after fixing $N_\mathrm{H,i}$ at the best-fitting
  value, which allows for easy tracking of the spectral evolution. The
  fits confirmed that the plateau spectrum had a smaller value of the
  photon index, thus is harder, than the decay spectrum at the
  confidence levels of 2.5$\sigma$ and 3.2$\sigma$ in XRT 030511 and
  XRT 110919, respectively. In XRT 170901, the photon index of the
  plateau spectrum is consistent with that of the decay spectra within
  $0.7\sigma$. We note that we could not rule out that the spectral
  softening might also be present in XRT 170901 but was missed due to the early termination of the observation before
  the flare ended.

  \subsection{The Long-term X-ray Variability}
  
Figure~\ref{fig:ltlc} plots the long-term X-ray flux curves of the
three new FXTs from all the \emph{Chandra}, \emph{XMM-Newton} and
\emph{Swift} observations (Table~\ref{tbl:obslog}). No extra flares
were detected. There was no significant persistent emission seen in any observation
in all the three XRTs either. Combining all available observations, we obtained the most stringent constraint on the
0.5--7.0 keV
persistent emission to be  $<$$6.3\times10^{-16}$ erg~s$^{-1}$~cm$^{-2}$,
$<$$1.0\times10^{-14}$ erg~s$^{-1}$~cm$^{-2}$, and
$<$$1.0\times10^{-15}$ erg~s$^{-1}$~cm$^{-2}$ ($3\sigma$ upper limits) for XRT 170901, XRT 030511,
and XRT 110919, respectively. 
Therefore, these FXTs have very large amplitudes, with  the peak to
the persistent X-ray flux ratio
$F_\mathrm{peak}/F_\mathrm{persistent}$$>$1320, $>$170, $>$430, respectively.  Because only one
flare was detected in each FXT, we estimated the upper limit of the
duty cycle $\gamma$ roughly as
$\lesssim1/T_\mathrm{tot}$, where $T_\mathrm{tot}$ is the total exposure time
of all the \emph{Chandra}, \emph{XMM-Newton} and
\emph{Swift} X-ray observations. We obtained $\gamma$ $\lesssim$$1.4\times10^{-3}$
ks$^{-1}$, $\lesssim$$1.4\times10^{-2}$
ks$^{-1}$,  and $\lesssim$$2.0\times10^{-3}$
ks$^{-1}$, respectively.

\subsection{The High-Energy Emission}
Our three FXTs were not associated with a GRB in the catalog compiled by Jochen Greiner (\verb|https://www.mpe.mpg.de/~jcg/grb110918A.html|).
The interplanetary network, which examines
high-energy data from a group of spacecraft equipped with $\gamma$-ray
burst detectors, such as \emph{INTEGRAL}, \emph{Swift}, and
\emph{Fermi}, provided a rough fluence upper limit of $10^{-6}$
erg cm$^{-2}$ for all the three FXTs (K. Hurley,  private
communication). 

There were Konus-Wind (KW) waiting mode
data for all the three FXTs, and no high-energy counterparts were found
from the $+/-$10 ks interval of each FXT (D. Svinkin, A. Ridnaia, D. Frederiks, private communication). Two upper limits were
calculated following \citet{kosvly2019} and \citet{risvfr2020}: an upper limit on the fluence for a burst lasting
less than 2.944 s and having a typical KW sGRB spectrum
(an exponentially cut off power law with $\alpha=-0.5$ and
$E_\mathrm{p}=500$ keV) and a limiting peak flux (2.944 s scale) for a typical lGRB spectrum
(the Band function with $\alpha=-1$, $\beta=-2.5$, and $E_\mathrm{p}=300$ keV).
Both were calculated for the interval +/-100 s relative to the start
time of each FXT and are in the 20--1500 keV band and at the 90\% confidence level.
The results are given in Table~\ref{tbl:properties}. The sGRB fluence
upper limit was about $5.5\times10^{-7}$ erg cm$^{-2}$ and the lGRB
peak flux upper limt was about $1.5\times10^{-7}$ erg cm$^{-2}$
s$^{-1}$ for all the FXTs. The fluence upper limit is a factor of
three more than the GRB 170817A \citep[$1.8\times10^{-7}$ erg
cm$^{-2}$;][]{govebu2017}. Therefore the KW limits can exclude
the association of our FXTs with bright GRBs, but fail to exclude the association with faint GRBs or orphan afterglows \citep{yaiona2002,ghsaca2015,lademo2017}.

\subsection{The Host Galaxies}
\label{sec:multiwav}

Figure~\ref{fig:hstxrt170901} shows the \emph{HST} images in four
broad bands, two in optical (F606W and F814W) and two in IR (F110W and
F160W), around the field of XRT 170901. The 95\% X-ray positional error of this FXT
overlaps with a highly elongated galaxy (``Gal1'' in
Figure~\ref{fig:hstxrt170901}). The chance probability to find XRT
170901 so close to a galaxy brighter than Gal1 in F160W (24.64 AB mag) was estimated
to be very low, only 0.5\%. Therefore we consider it as a
candidate host galaxy of XRT 170901. This galaxy might be irregular,
with the emission peaking to the west and slightly outside the 95\%
positional error circle of XRT 170901. The optical and IR colors of
this galaxy are
somewhat blue (Table~\ref{tbl:properties}), indicating a late-type galaxy with possibly strong
star-forming activity.  The other closest galaxy is ``Gal2'' in
Figure~\ref{fig:hstxrt170901}, $\sim$1 arcsec away from XRT
170901. Because it is well outside the 95\%
positional error circle of XRT 170901, it is very unlikely to host this
FXT. We note that Figure~\ref{fig:hstxrt170901} shows that there might
be a cluster of galaxies, with both Gal1 and Gal2 being its members.

Figure~\ref{fig:xrt030511optir} shows the Dark Energy Survey \citep[DES,][]{ababal2018} $r$-band and Gemini
$K_\mathrm{s}$-band images around the field of XRT 030511. There are
 DES and VISTA images in other filters around this FXT, but we saw no
clear host galaxy in any DES or VISTA image;
Table~\ref{tbl:properties} lists the detection limits of these images. In the very deep Gemini
$K_\mathrm{s}$-band image, there might be a very weak source (a
1\arcsec-aperture magnitude of 24.5$\pm$0.2 AB mag) within
the 95\% positional uncertainty of XRT 030511. Deeper images are
required to confirm this candidate counterpart.

Figure~\ref{fig:xrt110919optir} shows the DES $r$-band and VISTA
$K_\mathrm{s}$-band images around the field of XRT 110919, and no
clear host galaxy are seen in them. There are also DES and VISTA images in other filters  around the field of this FXT, without a clear host
galaxy seen in them either.
Table~\ref{tbl:properties} lists the detection limits of various
images. The ground-based images around the fields of XRT 030511
and XRT 110919 are not deep enough (e.g., $r>24.7$ AB mag, $J>22.0$ AB
mag) to rule out faint hosts as seen in
XRT 170901 and CDF-S XT2 ($m_\mathrm{F606W}=25.35$ AB mag,
$m_\mathrm{F160W}=23.85$ AB mag, Xue19).

\section{DISCUSSION}
\label{sec:discussion}

\subsection{Magnetar-powered X-ray transients}
\begin{figure}
\begin{center}
  \includegraphics[width=3.4in]{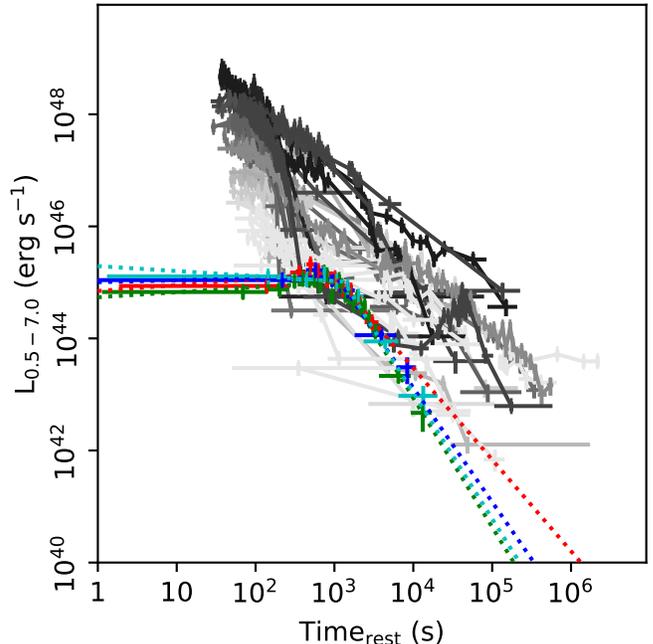}
\end{center}
\vskip -0.2in
\caption{X-ray luminosity curves for XRT 170901 (red), XRT 030511
  (green), XRT 110919 (blue), CDF-S XT2 (cyan), and sGRBs
  detected by \emph{Swift}/XRT and with redshifts (black/gray). The
  distances of our three FXTs were inferred from CDF-S XT2, assuming
  a constant value of $L_\mathrm{sd}t_\mathrm{sd}$. For clarity, the
  FXT data were rebinned to be $>4\sigma$ in each bin.
\label{fig:sgrbcmp}}
\end{figure}

We have shown that the three new FXTs that we discovered also from the
\emph{Chandra} archival data are similar to CDF-S XT2 in many aspects,
especially the flare light curve profile: a fast rise to a plateau
lasting 1--2 ks, followed by a steep decay of approximately
$t^{-2}$. Therefore these three new FXTs and CDF-S XT2 might have the
same origin. The intriguing explanation that Xue19 proposed for CDF-S
XT2, i.e., a magnetar-powered X-ray transient resulted from a BNS
merger, could then be applied to our three FXTs. In the case of CDF-S
XT2, the supporting evidence for the magnetar explanation
includes not only the unique light curve profile but also the host
properties. Xue19 found that the combined properties of the host
galaxy of CDF-S XT2 (stellar mass, star-formation rate, metallicity,
offset, and galaxy size) is much more commonly seen in sGRBs than in
lGRBs. The redshift and thus the distance of its host galaxy is known
($z=0.738$). Therefore Xue19 could obtain the peak X-ray luminosity of
CDF-S XT2 to be $3\times10^{45}$ erg s$^{-1}$, much lower than those
seen in the X-ray afterglows of sGRBs. Therefore CDF-S XT2 is
consistent with an off-axis jet configuration. For our FXTs, only the
host of XRT 170901 was detected, thanks to the deep \emph{HST}
images. Although its host might be a late-type galaxy of a modest
star-formation rate, given the blue \emph{HST} colors, this FXT is
most likely not in the star-forming region, with a significant offset
from the peak emission of the host. Therefore, this FXT is probably
also of a compact star merger origin. Our three FXTs had similar peak
fluxes as CDF-S XT2 ($\sim10^{-12}$ erg s$^{-1}$ cm$^{-2}$).  Therefore,
they would have similar peak luminosity and are also more likely in an
off-axis jet configuration if they are all at similar
redshifts as CDF-S XT2.

The X-ray light curves of lGRBs and sGRBs typically include an initial very steep
decay, which is most likely the tail
of the prompt emission, and the long-lasting afterglow, which is
caused by the forward shock from the interaction of the jet with the
external medium \citep{nokogr2006,be2014}. However, they often also show many complicated
features that can signify the presence of the magnetar central engine \citep[e.g.,][]{dawawu2006,be2014,be2015}:
a plateau phase, precursors, and flares. The plateau phase, following
the initial steep decay, is
very common in both lGRBs \citep[up to $\sim$80\%, e.g.,
][]{evbepa2009,mazabe2013,mecoro2014} and sGRBs \citep[$\sim$50\%,
e.g., ][]{roobme2013,dasabe2014}. The
main argument to attribute this phase to the spinning-down energy of a
newly born magnetar instead of a forward shock is that the plateau
in some cases in both lGRBs and sGRBs was followed by a sudden drop
\citep{roobme2013}, which is hard to explain with a forward shock
model but can be naturally
explained by the collapse of a supramassive magnetar to a BH
\citep{lyobzh2010, roobme2013}. The magnetic field strength and the
spin period inferred from the magnetar model for the plateaus seen in
lGRBs and sGRBs are around $B\sim10^{15}$ G and $P\sim1$ ms
\citep{rogoda2014}.

Our three FXTs and CDF-S XT2 are most similar to the plateau component
in the X-ray light curves of GRBs. An alternative explanation for the
low levels of the peak X-ray flux and the high-energy emission in
these FXTs, compared with typical GRBs, is that they are distant GRBs,
but this is very unlikely because an initial steep component
associated with the prompt emission was not present in at least three
of these four FXTs (except XRT 110919) whose early phases were fully covered by the
X-ray observations.

If all these four FXTs are magnetar-powered, then there is a question of
whether the newly born magnetars in these systems are stable or
not. Because their decays were all consistent with $t^{-2}$ within $2\sigma$, as
expected for the spindown luminosity of a stable NS, 
it is more likely that a stable NS was formed in these systems
\citep[see however][]{luyula2021}.

There is a correlation between the spindown luminosity and the
spindown time scale: $L_\mathrm{sd}t_\mathrm{sd}\propto
P^{-2}_\mathrm{0,-3}$, where $P_\mathrm{0,-3}$ is the initial spin
period in millisecond \citep{zhang2013}. Then we can infer the redshift of 
0.60, 0.39, and $<$0.82 and
the luminosity distance of 3.6 Gpc, 2.14 Gpc, and  $<$5.2 Gpc, for XRT
170901, XRT 030511, and XRT 110919, respectively, if we assumed that
the magnetars in them
have the same
initial spin periods as in CDF-S XT2. The upper limit was obtained for
XRT 110919 because we only obtained the lower limit on the spindown
time scale (the start of this FXT was missed by the observation).

Based on these distances, we obtain the luminosity curves in Figure~\ref{fig:sgrbcmp} for these
magnetar-powered FXT candidates. They look remarkably similar to
each other. Such a similarity indicates that the beginning of
  XRT 110919 might not be missed
  much by the observation. The plot also includes sGRBs that have redshifts
 and were
followed up by the XRT obtained from the Swift Burst Analyser
\citep{evwios2010}. It shows that if the distances that we inferred
for our FXTs are correct, their initial luminosities would be much
lower than seen in sGRBs, which would suggest the lack of the prompt
emission and thus indicate an off-axis jet configuration.

We followed \citet{batrsc2017} to have a simple estimate of the event
rate for these magnetar-powered FXT candidates.  Because we considered only relatively bright FXTs, we assumed
that they can be detected and identified in any CCD of any ACIS
observation longer than $>$5 ks. With four magnetar-powered FXTs
discovered in our search (including CDF-S XT2), we obtained a rate of $4.5^{+3.6}_{-2.1}$ events
deg$^{-2}$ yr$^{-1}$ \citep[the error is 1$\sigma$,
following][]{ge1986}. To convert it to a local volumetric rate density, we
assumed the maximum redshift that the observations can reach is 1.0
and we obtained a rate density of $150^{+120}_{-70}$ Gpc$^{-3}$
yr$^{-1}$. This is roughly consistent with that obtained by
\citet{batrsc2017} for ``CDF-S XT1''-like events and that by Xue19 for
``CDF-S XT2''-like events and is also broadly consistent with the BNS merger event rate
density \citep[$1.5^{+3.2}_{-1.2}\times10^{3}$ Gpc$^{-3}$
yr$^{-1}$,][]{ababab2017c}, given the large uncertainties of these estimates.

A definite signature for the BNS merger nature of these FXTs would be
the simultaneous detection of both GW events and such kind of FXTs. We
consider whether such a source could be detected by the Burst Alert
Telescope (BAT) onboard \emph{Swift}. The BAT can reach a $5\sigma$
sensitivity (14--195 keV) of $2.9\times10^{-8}\sqrt{T/1\mathrm{s}}$
erg s$^{-1}$ cm$^{-2}$, where $T$ is the integration time
\citep{batuma2013}. The duration of the plateau phase in the rest
frame is about 1000 s for all these FXTs if they are at the redshifts
inferred above, so assuming an integration time of $T=1000$
s, the BAT can reach a sensitivity of $1.0\times10^{-9}$ erg s$^{-1}$
cm$^{-2}$.  The Advanced Ligo/Virgo can detect BNS mergers up to
around 150 Mpc currently \citep{ababab2020}. At this distance, CDF-S
XT2 would have a peak flux around $1.0\times10^{-9}$ erg s$^{-1}$
cm$^{-2}$ in the soft X-rays. Assuming a photon index of 2.0 (all
our three FXTs and CDF-S XT2 have photon index in the plateau phase
consistent with this value), these FXTs would have a BAT flux of also
around $1\times10^{-9}$ erg s$^{-1}$ cm$^{-2}$, which can be
marginally detected if it is in the BAT field of the view. The BAT was
not pointed toward GW170817 \citep{evceke2017} or the second strong BNS GW
event candidate GW190425 \citep[at a distance of $159^{+69}_{-71}$
Mpc;][]{sabali2019} at the times when these GW events were triggered.

The XRT onboard \emph{Swift} has also carried out a fast scan
 over the error regions of GW events
that are good BNS merger candidates in order to search for the EM
counterparts \citep{evkepa2016}. The observations typically have an
exposure time of 70 s each or sensitivity $\sim1.0\times10^{-11}$ erg
s$^{-1}$ cm$^{-2}$. Assuming a distance of $150$ Mpc for the GW
events, our FXTs and CDF-S XT2 can stay above the detection limit
of these XRT scans ($3\times10^{43}$ erg s$^{-1}$) for up to $\sim$5 ks (rest-frame). Therefore, the XRT scans need to
be carried out very quickly in order to detect magnetar-powered FXT signals from the BNS GW events. The XRT observed the EM counterpart to
GW170817 at 0.62 days after the GW trigger, with the $3\sigma$ upper
limit of $5\times10^{40}$ erg s$^{-1}$ \citep{evceke2017}. Based on
Figure~\ref{fig:sgrbcmp}, our three FXTs and CDF-S XT2 can
still be detected at the luminosity around $1.0\times10^{41}$ erg
s$^{-1}$ at 0.62 days, marginally above the detection limit. The non-detection could be either because the
magnetar had collapsed by 0.62 days, or the spectra were softened
significantly and shifted out the X-ray bands, or a magnetar was simply not
formed in this GW event.

\subsection{Alternative Explanations}
\begin{figure}
\begin{center}
  \includegraphics[width=3.4in]{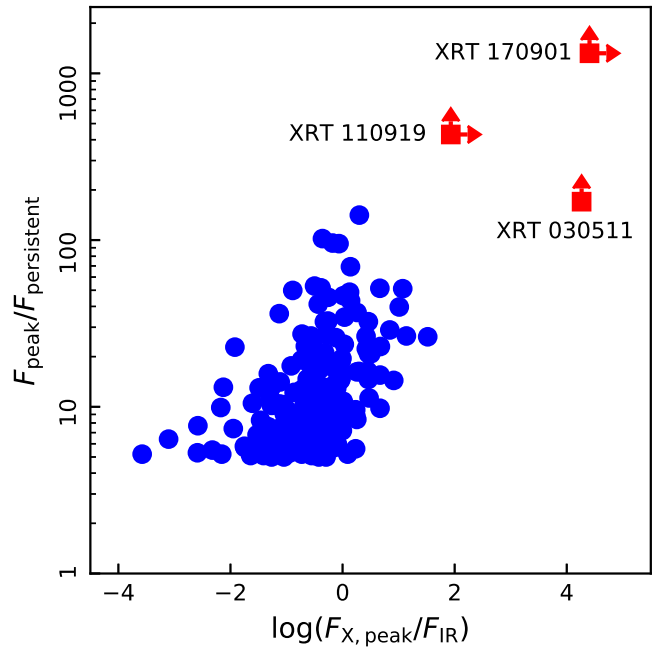}
\end{center}
\vskip -0.2in
\caption{The flare peak to persistent X-ray flux ratio versus the
  X-ray (0.2--12 keV, peak) to IR ($K_\mathrm{S}$-band, or F160W for XRT
170901) flux ratio for
  stellar flares \citep{liweba2012} and our three FXTs. Only the stellar
  flares of the  peak to persistent X-ray flux ratio above 5 were
  selected in \citet{liweba2012}.
\label{fig:starscomp}}
\end{figure}
An alternative explanation for our three FXTs and CDF-S XT2 is the
shock breakout (SBO) of a core-collapse supernova
\citep{waka2017,lena2020}.  SBOs are expected to produce short-lived
X-ray bursts lasting for a few minutes. The best known example of  SBOs
is SN2008D \citep{sobepa2008}, which lasted for about 400 s and
exhibited a fast-rise-and-exponential-decay profile. Dedicated search
over \emph{XMM-Newton} archival data found a dozen SBO candidates
\citep{noesti2020,alla2020}. With very fast rise and a plateau in the
X-ray light curves, our FXTs and
CDF-S XT2 are distinct from these SBO candidates and should belong to
a different class of objects.

\citet{peyash2019} considered CDF-S XT2 and CDF-S
XT1 as accretion-driven flares from tidal disruption of white dwarfs
by intermediate-mass BHs. For CDF-S XT2,
\citet{peyash2019} inferred a viscous accretion time scale of 1.5 ks,
which is rather long and requires a radiatively efficient and
geometrically thin disk. However, CDF-S XT2 appeared highly
super-Eddington at the peak, which then requires the presence of a significant beaming effect (a beaming factor of more than 1000) in the
system, possibly in the form of a jet. Although it is hard to
completely rule it out, we think this is very unlikely, given that we
have discovered three more similar systems.

Finally, our three new FXTs are unlikely due to stellar
flares. Figure~\ref{fig:starscomp} compares our FXTs with the
stellar flares discovered in \citet{liweba2012} from the
\emph{XMM-Newton} catalog. These flares mostly have 
$F_\mathrm{peak}/F_\mathrm{persistent}<150$ and the X-ray (peak) to IR flux
ratio $F_\mathrm{X, peak}/F_\mathrm{IR}<1.6$, while our FXTs have much
higher amplitudes ($>$170--1320) and
much higher peak X-ray flux relative to the persistent IR flux, with $F_\mathrm{X, peak}/F_\mathrm{IR}>4.5$, $\approx4.3$, $>1.9$, for XRT
170901 (the presumptive stellar IR flux was estimated from the
detection limit due to the lack of a point-like counterpart in the
\emph{HST} images), XRT 030511 and XRT 110919, respectively. Besides, if they
are stellar flares, they would have to be very far away,  $>$31 kpc, $\approx15$ kpc,
and $>$5 kpc for  XRT
170901,  XRT 030511, and XRT 110919 even assuming a very faint, late-type star of
M6 \citep{coivsc2007}. This is very unlikely given their very high
Galactic latitudes of $b=-69^\circ$, $-35$\degr, and $-84$\degr,
respectively.

\section{CONCLUSIONS}
\label{sec:conclusions}
We have discovered three FXTs from the \emph{Chandra} archival data. They share several similar properties.
\begin{itemize}
\item The flares have a large amplitude, with the peak flux a factor of
          $>$170--1320 above the persistent level.
\item The rise phase was not clearly detected and should be
          very short ($\lesssim42$ s for XRT 170901 and $\lesssim9$ s
          for XRT 110919; the rise of XRT 030511 was most likely missed by the observation).
\item The flare light curve profiles can be fitted with a broken
            powerlaw of the initial index consistent with $0$ and the second index
            $\sim-2$. The initial plateau last $\sim$1.5--1.9 ks.
\item The X-ray spectra were hard in the plateau phase in
              all the three FXTs and seemed to soften in the
              decay at least in two of them whose decay phase was well
              covered (XRT 030511 and XRT
              110919). The spectral softening is not obvious in XRT
              170901, whose late decay phase was missed by the observation.
\item Persistent X-ray emission was not detected, with upper
          limits of $<1.0\times10^{-15}$ erg~s$^{-1}$~cm$^{-2}$ in XRT
          170901 and XRT 110919 and $1.0\times10^{-14}$ erg~s$^{-1}$~cm$^{-2}$ in XRT 030511.
\item Only a single flare was detected in all the three FXTs,
          though they are covered in many other observations. Therefore,
          these FXTs might be non-recurrent.
\item No GRBs were found to be associated with these FXTs.

\item The host galaxies are very faint, based on the
            detection of the host for XRT 170901 and the detection
            limits for XRT 030511 and XRT 110919.
\end{itemize}

These properties are similar to those seen in CDF-S XT2, which is a
strong candidate for a magnetar-powered transient from a BNS
merger. Such a magnetar-powered FXT signal could be detected by the
\emph{Swift}/BAT if it happens to point toward the GW events or by the
\emph{Swift}/XRT scans if they can be carried out quickly, optimally
within 5 ks. We argue that these FXTs are very unlikely SBOs, tidal
disruption events, or
          stellar flares.

\acknowledgments
This work is supported by NASA ADAP Grant
NNX10AE15G. We thank Raffaella Margutti for providing the list of
sGRBs that have redshifts.

\end{document}